\documentstyle[aaspp4,psfig,amstex]{article}

\def\ic{IC\thinspace{443}}
\def\ts{\thinspace}

\begin{document}
 
\title{\large \bf A Bow-Shock Nebula Around a Compact X-ray Source \\ in
  the Supernova Remnant IC\thinspace{443}}

\author{
Charles M. Olbert\altaffilmark{1},
Christopher R. Clearfield\altaffilmark{1},
Nikolas E.  Williams\altaffilmark{1},
Jonathan W. Keohane\altaffilmark{1}, \\
Dale A. Frail\altaffilmark{2}}

\altaffiltext{1}{North Carolina School of Science and Mathematics,
  1219 Broad St., Durham, NC 27705}

\altaffiltext{2}{National Radio Astronomy Observatory, P.~O.~Box O,
  Socorro, NM 87801}

\centerline{Accepted to the Astrophysical Journal Letters as of March 15$^{th}$, 2001.}

\begin{abstract}
  We present spectra and high resolution images of the hard X-ray feature along
  the southern edge of the supernova remnant IC\,443. Data from the
  {\it Chandra X-ray Observatory} reveal a comet-shaped nebula of hard
  emission, which contains a softer point source at its apex.  We also
  present $\lambda$20\,cm, $\lambda$6\,cm, and $\lambda$3.5\,cm images
  from the {\it Very Large Array} that clearly show the cometary
  nebula.  Based on the radio and X-ray morphology and spectrum,
  and the radio polarization properties, we argue that this object is
  a synchrotron nebula powered by the compact source that is physically
  associated with IC\,443.  The spectrum of the soft point source is adequately
  but not uniquely fit by a black body model (kT=0.71$\pm$0.08\,keV, 
  L=$(6.5\pm0.9)\times10^{31}$\,erg\,s$^{-1}$).  The cometary
  morphology of the nebula is the result of the supersonic motion of
  the neutron star (V$_{\rm{NS}}\simeq250\pm$50 km\,s$^{-1}$),
  which causes the relativistic wind of the pulsar to terminate in a
  bow shock and trail behind as a synchrotron tail.  This velocity is
  consistent with an age of 30,000 years for the SNR and its
  associated neutron star.
\end{abstract}


%\clearpage

\section{Introduction}\label{sec:intro}

The mixed-morphology Galactic supernova remnant \ic\ (l,b =
189.1\arcdeg, +3.0\arcdeg) has been the subject of extensive studies
at all wavelength bands (\cite{fes84}; \cite{bs86}; \cite{vjp93};
\cite{aa94}).  \ic\ is especially well known for its clear interaction
with surrounding molecular clouds (\cite{bhhe90}; \cite{bb00}).  With
its large variety of shocked molecular species detected along its SE
edge, \ic\ has become a standard laboratory for studying shock
chemistry (van Dishoeck {\it et al.} 1993 and references therein).

\ic\ is also of interest because it is coincident with the
unidentified EGRET source 2EG\,J0618+2234 (\cite{sd95};
\cite{ehks96}), and thus has stimulated a good deal of theoretical
work aiming to explain the production of GeV $\gamma$-rays by shell SNRs
(\cite{sdm96}; \cite{ssdm97}; \cite{gps98}; \cite{bergg99};
\cite{bceu00}).

In the X-ray band, Petre et al.~(1988)\nocite{pss+88} found that the
bulk of the emission was thermal in origin with a temperature
$\sim$${10}^7$ K, typical of middle-aged SNRs. However, Wang et al.\ 
(1992)\nocite{wahk92} found evidence in the {\it Ginga} satellite data
for a hard X-ray component in the spectrum of \ic\ extending up to 20
keV\@.  Follow-up observations with the {\it ASCA} satellite revealed
that the bulk of this hard X-ray emission came from a single,
unresolved feature located along the edge of the radio shell where the
SNR/molecular cloud interaction was strongest (\cite{kpg+97},
hereafter K97)\@.  Furthermore, the hard X-ray feature is positionally
coincident with a region where the radio spectral index is
considerably flatter than the SNR as a whole (i.e $\alpha<$ 0.24 vs
$\alpha$ = 0.42, where $F_\nu\propto\nu^{-\alpha}$) (\cite{gre86};
\cite{kpu94}).

K97 considered different explanations for the origin of this hard
X-ray emission, and favored a model in which synchrotron emission
was being produced by TeV electrons in a region of enhanced particle acceleration,
resulting from the SNR/molecular cloud interaction. Recent BeppoSAX data 
analyzed by Bocchino \& Bykov (2000)\nocite{bb00} also support
the K97 interaction model. An equally viable model, favored by
Chevalier (1999, hereafter C99)\nocite{che99}, posits that the hard
X-ray feature is synchrotron emission powered by an energetic neutron
star. To distinguish between these two competing hypotheses we have
undertaken high resolution X-ray and radio observations toward the
hard X-ray feature. These new Chandra and VLA observations strongly
favor C99's neutron star hypothesis.

\section{Observations and Analysis}\label{sec:anal}

\subsection{Chandra X-Ray Observatory}

\begin{figure}[tb]
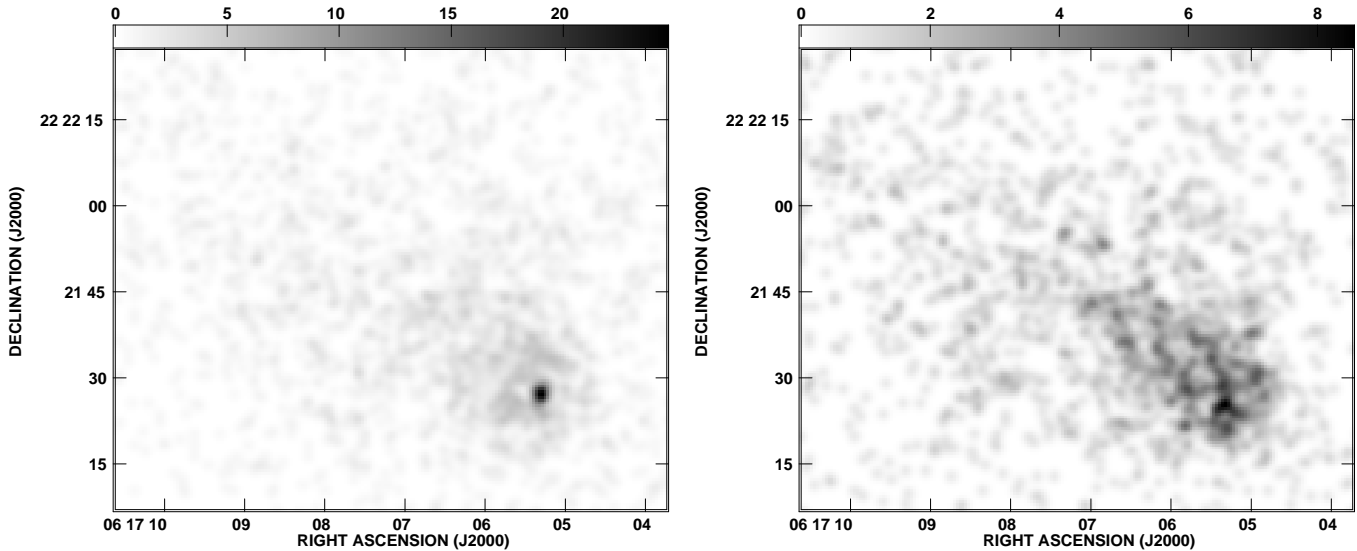
 
  \centerline{\hbox{\psfig{figure=xsoft.ps,angle=0,width=9cm}}
{\psfig{figure=xhard.ps
,angle=0,width=9cm}}}
\caption[]{X-ray intensity images of the hard X-ray feature on the
  southern edge of \ic. The image on the left is in the soft
  ($E < 2.1$\,keV) photon energy band, scaled between 0 to 24.6
  counts per beam. The image on the right is in the hard ($E >
  2.1$\,keV) photon energy band, scaled between 0 and 8.5 counts per
  beam. Both of these Chandra images have been smoothed with a
  2\arcsec\ beam. The cometary nebula is most apparent in the hard
  band image (right), while the soft point source stands out well
  against the surrounding nebula in the soft band image (left).
\label{fig:xray}}
\end{figure}

The {\it Chandra X-Ray Observatory} performed a short (10 ks)
observation of IC\,443 on April 10, 2000 during the first cycle of
guest observations (A01).  The hard X-ray feature was centered on the
I3 chip of the Advanced CCD Imaging Spectrometer (ACIS). To produce
Chandra event files, we followed the standard CIAO procedures outlined
in Version 1.3 of the {\it CIAO} Beginner's Guide for {\it CIAO}
Release 1.1, using the {\tt acisD2000-01gainN0001.fits} gain file.
From the events file, we extracted high resolution images in both hard
($E > 2.1$\,keV) and soft ($E < 2.1$\,keV) spectral
bands.  Smoothed versions of these images are presented in
Figure\,\ref{fig:xray}.

The hard X-ray source is resolved by this high resolution
($\sim$1\arcsec) Chandra observation (Figure~\ref{fig:xray}) as a
nebular region of diffuse emission with a cometary tail.  Within the nebula 
there is an unresolved point source that we have designated as
{\em CXOU\,J061705.3+222127}\@. The nebula of hard emission exhibits 
bow-shock morphology with a width of $\simeq$ 35$\arcsec$ and a 
minimum standoff distance of $r_{s}$ $\simeq$ 8.5$\arcsec$.
The X-ray point source lies at the apex of the nebula, located at
(epoch 2000) $\alpha$~=~06$^h$17$^m$05.31$^s$, $\delta$~=~22\arcdeg
21\arcmin 27\arcsec.3, with errors of $\pm$2\arcsec\ in each
coordinate.  It is especially visible in the soft
($E<2.1$\,keV) X-ray band (Figure\,\ref{fig:xray}). 

Likewise, in order to compare the X-ray luminosity of the cometary nebula with
statistical studies of pulsar wind nebulae (PWNe) 
(e.g.~Seward \& Wang 1988\nocite{sw88}, see \S{\ref{sec:dis}}), we 
integrated the power-law model of K97 over the {\it Einstein} 
band (0.2-4~keV) -- thus deriving 
L$_x\sim5\times{10}^{33}\,d^2\,(1.5 {\rm\,kpc})^{-2}\,erg\,s^{-1}$\@. 
Inherent in this L$_x$ calculation is the assumption that the cometary
nebula is dominated in soft X-rays by the same power-law exibited at hard 
energies with ASCA\@.  The Chandra flux (over the same range) yields 
consistent results, albeit with larger uncertainties.

\subsection{Very Large Array}

All radio observations were made on August 26 and December 31, 1997
with the {\it Very Large Array} (VLA\footnotemark\footnotetext{The VLA is
operated by the National Radio Astronomy Observatory, a facility of the
National Science Foundation.}) in the C and D arrays, respectively. A log of the
observations is summarized in Table \ref{VLA}. The data acquisition
and calibration were standard, using J0632+103 as a phase calibrator
and 3C\ts{138} as both the flux density and polarization angle
calibrator.

\begin{figure}[tb]
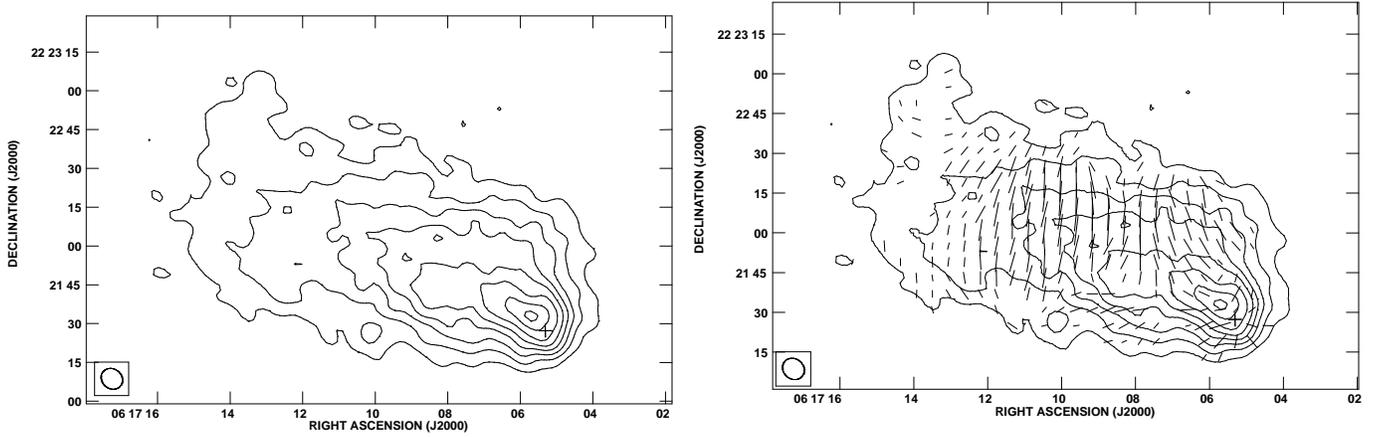

  \centerline{\hbox{\psfig{figure=ic443_pwn_xi.ps,angle=270,width=9cm}}
{\psfig{figure=ic443_pwn_xp.ps,angle=270,width=9cm}}}
\caption[]{Total intensity image of the cometary nebula in \ic\
  (left) at a frequency of 8.46 GHz. Contours are in steps of 0.5 mJy
  beam$^{-1}$. The size of the
  synthesized beam is shown in the lower left corner.
  The same total intensity
  image with vectors indicating the magnitude and direction of the polarized 
intensity (right).  The scale is 50\,$\mu$Jy\,beam$^{-1}$\,arcsec$^{-1}$\@. 
The cross marks the position of CXOU\,J061705.3+222127\@.
\label{fig:radioip}}
\end{figure}

The cometary and bow-shock morphologies of the extended source are even 
more pronounced at radio wavelengths, extending some 2\arcmin\ to the 
northeast of the soft X-ray source.  Although we display only the 
$\lambda$3.5\,cm image in Figure \ref{fig:radioip}, similar structure is visible 
at $\lambda$6\,cm and $\lambda$20\,cm.  The peak in the radio emission
lies about 6\arcsec\ to the northeast of the compact X-ray source.
There is no radio point source ($>2$ mJy) coincident with the X-ray
point source. The distribution of linearly polarized emission is also
shown in Figure \ref{fig:radioip}.  The direction of the E-vectors are
generally parallel to the shock normal, as would be expected for
evolved SNRs with circumferential magnetic fields.  The character of
the magnetic field evidently changes substantially in the area
dominated by the X-ray emission.  Here the magnetic field direction
appears to wrap around the ``head'' of the hard nebula, tracing a
bow-shock morphology. The average degree of polarization in this
region is 8\%. This is likely a lower limit since, owing to the
compact size of the nebula, there is likely some beam depolarization
occurring. To the northeast, where the field is more ordered, the
percentage polarization is uncharacteristically strong for an evolved SNR, 
exceeding 25\% in several locations.

The integrated flux density over the entire length of the cometary
nebula is given for three frequencies in Table \ref{VLA}. An
approximate value of 230$\pm$200 mJy was also obtained from the 327
MHz image presented by Claussen et al.~(1997\nocite{cfg+97}). The
increased error in estimating the flux density at low frequencies is
due to the uncertainty in subtracting out the background shell
emission from the SNR. Within the errors, the radio spectrum is well
represented by a flat spectral index $\alpha_R$ $\simeq$ 0.0 and a mean
flux density of 206 mJy. In contrast, K97 measured an X-ray spectral
index of $\alpha_X\simeq$1.3$\pm$0.2, which implies a break in
the spectrum near 4$\times${10}$^{13}$ Hz.  The radio luminosity of
the cometary nebula, determined by integrating the spectrum from 10
MHz to 100 GHz, is L$_R$=5.5$\times${10}$^{31}
\left(d/1.5{\rm kpc}\right)^2$ erg\,s$^{-1}$, or less than 1\%
of L$_R$ for \ic\ as a whole. 

\begin{deluxetable}{rrll}
\tabcolsep0in\footnotesize
\tabcolsep0.3in
\tablewidth{\hsize}
\tablecaption{Summary of VLA Observations of the Cometary Nebula \label{VLA}}
\tablehead
{
        \colhead{Frequency} & 
        \colhead{Time} & 
        \colhead{Beam} & 
        \colhead{F$_\nu$}  \\
        \colhead{(GHz)} & 
        \colhead{(min)} & 
        \colhead{$(\arcsec \times \arcsec)$} & 
        \colhead{(mJy)}
}
\startdata
8.46  &  13 & 8.6  $\times$ 7.6   & 195 $\pm$ 8  \\
4.86  &  36 & 5.0  $\times$ 4.8   & 173 $\pm$ 11\\
1.46  &  53 & 15.5 $\times$ 14.5  & 229 $\pm$ 34
\enddata
\tablecomments{\footnotesize{The columns are (left to right), (1)
    Observing frequency, (2) total time on source, (3) angular
    resolution, and (4) integrated flux density of the cometary
    nebula.}}
\end{deluxetable}

\subsection{Spectra}

In addition to the spectral properties of the hard nebula, a spectrum 
of CXOU\,J061705.3+222127 was extracted from this short {\em Chandra} 
observation.  The surrounding hard nebula was subtracted as 
background (i.e. an annulus with an inner and outer radius of 4.3\arcsec\ 
and 23.6\arcsec\ respectively).  These spectra were binned by a factor of 
256 (PHA channels), and the softest three ACIS-I channels were ignored.  
As discussed above, the point source is significantly softer than the 
surrounding hard nebula, implying a likely thermal origin for the emission.
For this reason, we modeled the spectrum as black body radiation, though
this fit was not statistically unique due to large uncertainties in the small
data set.

The best-fit ($\chi _{\nu} ^2 \sim 1.1$) black body temperature 
and luminosity are kT=0.71$\pm$0.08~keV
and L=$(6.5\pm0.9)\times10^{31}$\,erg\,s$^{-1}$
(i.e. A=$0.025\pm0.003$($d1.5~{\rm kpc})^{2}$~km$^2$).
The 1-5\,keV X-ray flux of the point source is $\sim 2 
\times 10^{-13}$ erg\,s$^{-1}$ cm$^{-2}$.  This fits contained 
two free parameters by assuming K97's best-fit column density 
(N$_{\rm H}$=1.3$\times \! 10^{21}$cm$^{-2}$)\@. 

\begin{figure}[tb]
  \centerline{\hbox{\psfig{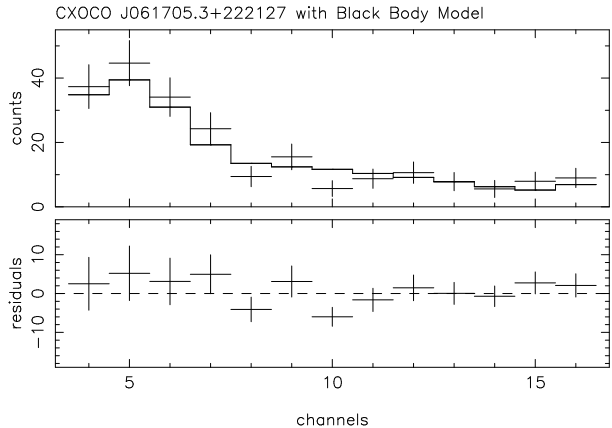}}}
\caption[]{The background subtracted spectrum of the soft point source, 
  fit with a black body emission model (kT=0.7$\pm$0.1~keV and 
  A=$0.02\pm0.01(d/1.5~{\rm kpc})^{2}~\rm{km}^2$), assuming a 
  previously measured column density.  Channels 5, 10 and 15 
  correspond to photon energies of 1.5, 2.9, and 4.2 keV respectively. }  
\label{fig:spectrum}
\end{figure}

\section{Discussion}\label{sec:dis}

The soft X-ray point source in IC\,443 is one of a growing number of
young neutron star candidates associated with supernova
remnants (\cite{hel98}). These neutron stars have a diverse range of
properties, suggesting that they do not originate from a homogeneous
population (\cite{cph+00}). The long-period (8-10 s) and claimed
high-field (10$^{14}$-10$^{15}$\,G) of anomalous X-ray pulsars (AXP) and
soft gamma-ray repeaters (SGR), differ from the canonical young radio
pulsars with periods P$\sim$15-500 ms and field strengths of 
B$\sim${10}$^{12}$-10$^{13}$\,G\@. The radio quiet neutron stars (RQNs),
with their thermal X-ray emission and lack of radio pulsations, may be
altogether different from either of these two classes of objects
(\cite{pza+00}).

In the absence of detectable pulsations, canonical radio pulsars can be distinguished
from AXPs, SGRs, and RQNs by the presence of an extended synchrotron
nebula, powered by the energetic relativistic wind (\cite{gbs00}).
It is argued that the cometary nebula around the X-ray point source in 
IC\,443 is most likely such a PWN. This nebula has all the
expected observational characteristics (C99\nocite{che99};
\cite{gsf+00}): flat spectrum radio emission ($\alpha_R$ = 0.1 to
0.3), a steep X-ray spectrum ($\alpha_X$ = 1.0 to 1.5) and a high
degree of linear polarization ($>5\%$). Morphologically it closely
resembles PWN detected toward the SNRs W\ts{44} and
G\ts{5.4$-1.2$} (\cite{fk91}; \cite{fgg+96}), both of which contain
active pulsars. It has been argued (\cite{fk91}; \cite{fgg+96}) that these 
nebulae have developed bow shocks as a result of ram pressure 
confinement of the pulsar wind, due to the high space velocity of their 
pulsar through the surrounding medium.  We assert that the most 
tenable explanation for the X-ray point source in IC\,443 is that it is a 
young pulsar which has traveled (ballistically) from its birth place to 
its present location, near the edge of the decelerating SNR.

Accepting this hypothesis we proceed with inferring the salient
physical properties of the neutron star and corresponding PWN.
Constraints can be placed on the velocity of the neutron star by a
variety of methods. The offset of the object from its origin gives a
measure of its transverse velocity V$_{\rm{NS}}$.  Measuring this
offset in the case of IC\,443 is made difficult owing to the uncertain
location of the explosion center (\cite{aa94}; \cite{cfg+97}).  We 
note that the neutron star has traveled
a fraction $\beta\simeq${0.9} of this projected distance (approximated from 
the distance between the soft point source and the shock), giving
V$_{\rm{NS}}=\beta$r$_s/t_s$. Likewise, the shock velocity of the SNR
can be similarly expressed as V$_s=c_\circ$r$_s/t_s$, where
$c_\circ$ is a constant equal to 2/5 for a remnant in the
Sedov stage, while C99\nocite{che99} argues for a value of 
$c_\circ=3/10$ for IC\,443 based off of the parameters of his
model.  Combining both equations for
V$_{\rm{NS}}$ and V$_s$ together, and using the value of V$_s=100$ 
km\,s$^{-1}$ favored by C99\nocite{che99}, gives
V$_{\rm{NS}}=V_s\,\beta{c_\circ}^{-1}$=225-300 km s$^{-1}$, a
result that is independent of distance and only weakly dependent on
the evolutionary state of the SNR (i.e. ${c_\circ}$).

It should be noted that the tail of the hard nebula does not point towards
the geometric center of the SNR.  However, the blast center and geometric
center of an SNR can be quite different in the presence of large-scale density 
gradients (\cite{dpj96}; \cite{hp99}).  Such a density gradient, combined with
the potential of a cross-wind, could possibly account for the apparent discrepancy.

We are able to make a second approximation of V$_{\rm{NS}}$ with
respect to the local medium by means of the nebula's bow shock
morphology.  K97 fit the SNR with an X-ray temperature kT $\simeq$ 1
keV, implying a sound speed c$_{local}\simeq$100 km\,s$^{-1}$.  An
approximation of the bow shock angle of 90\arcdeg-120\arcdeg\ gives a
minimum Mach number M$\simeq$1.1-1.5 through a standard geometric
calculation.  A fundamental determination of the speed of sound in this
particular region of the SNR gives velocity with respect to the local 
medium of 
V$_{\rm{NS\,local}}$=150\,M$_{1.5}$\,{kT$_{1\,\rm{keV}}^{0.5}$ km\,s$^{-1}$}.  
Accounting for
radial expansion of local plasma ($\le$100 km\,s$^{-1}$, this velocity 
could be as high as 250 km\,s$^{-1}$.  Both methods of determining 
V$_{\rm{NS}}$ are in good agreement with a model by C99\nocite{che99}, 
indicating a remnant age on the order of 30,000 years (i.e. $\tau_{30}\simeq$1).

The energy source for the synchrotron nebula must ultimately be
derived from the particles and field generated by the compact central
neutron star and therefore some fraction of the its spindown
power \.E is powering this emission.  Empirical relations have been
derived for known PWN between \.E and the observed X-ray and radio luminosities
(L$_x$ and L$_R$) that enable us to estimate the rotation period P and the
magnetic field B of the compact object. 

 For the value of L$_x$ derived in \S{\ref{sec:anal}} we obtain 
\.E$\sim$2$\times$${10}^{36}$\,erg\,s$^{-1}$ (Seward and Wang 1988)\@.
\nocite{sw88} Frail \& Scharringhausen~(1997)\nocite{fs97} and 
Gaensler~et~al.~(2000)\nocite{gsf+00} note
that \.E$\sim$10$^{4}$L$_R$ for known radio PWN, which yields a
value of \.E$\sim$6$\times$${10}^{35}$, so in the discussion that 
follows we adopt \.E=10$^{36}$\,ergs\,s$^{-1}$ 
(i.e.~\.E$_{36}$=1).  Equating the pressure of
the wind \.E/$4\pi{c}r_w^2$, to the ram pressure due to the neutron star's
motion through the surrounding gas $\rho$\ts{V}$_{\rm PSR}^{2}$, we
derive a density of $\sim$0.1 cm$^{-3}$ for our adopted values of \.E
and {V}$_{\rm NS}$. This density is more typical of that expected for
the hot X-ray interior of the SNR, and is orders of magnitude below
that expected in the dense molecular ring against which the NS and
its nebula are projected against (see van Dishoeck et al.~1993).  

The spindown luminosity of a pulsar is related to its current period P
and dipolar surface field B by \.E\,$\propto$\,B$^2$P$^{-4}$, while the
characteristic age of the pulsar is expressed as $\tau_c$\,$\propto$\,B$^{-2}$P$^{2}$\@. 
It is straightforward to show that for a pulsar born
spinning rapidly (i.e. P$_\circ$$<<$P), which loses energy
predominately via magnetic dipole radiation, 
P=145\,ms\ts($\tau_{30}$\.E$_{36})^{-1/2}$ and 
B=3.3$\times${10}$^{12}$\,G\,\ts$\tau_{30}^{-1}$\.E$_{36}^{-1/2}$.
Thus, with our nominal choice of parameters, the period and magnetic 
field closely match those of other young pulsars 
(e.g. PSR B1757$-24$; \cite{fk91}).

In the light of this result, it is worth re-examining the nature of
the unidentified EGRET source 2EG~J0618+2234 
(l,b=189.13\arcdeg,+3.19\arcdeg) (\cite{sd95}; \cite{ehks96}) toward \ic\@.  
Nel et~al.~(1996)\nocite{n96} note that all known gamma-ray pulsars 
have a large (greater than 0.5) ratio of ${\dot{E}}_{33}/{d_{kpc}}^{2}$.  
For our derived $\dot{E}\sim10^{36}$\,erg\,s$^{-1}$ and d$\sim$1.5\,kpc,
this ratio is over 400, similar to that of the gamma-ray emitting 
PSR~B1706$-$44, strongly suggesting that the
unidentified EGRET emission originates from the neutron star or the PWN.  The
difficulty with this interpretation is that our proposed neutron star and PWN
lie several arcminutes outside the formal 95\% error radius of the
EGRET source.  However, since the position of EGRET sources in the
Galactic plane are subject to systematic error (e.g. \cite{h97}), it
is entirely possible that this object may be the source of
this gamma-ray emission.

\section{Conclusion}

In summary, the hard X-ray feature in \ic\ is best interpreted as a
wind nebula powered by a young neutron star. The PWN interpretation is
suggested first on morphological grounds by the Chandra and VLA
images, which show a comet-shaped nebula with a soft X-ray point
source at its apex.  The measurement of significant polarization and a
flat radio spectrum further strengthens the argument that this is
synchrotron emission from a PWN\@. 
%%Maybe remove this sentence
The X-ray spectrum of the point
source is appropriately, but not uniquely fit by a black body model,
suggesting that the emission is thermal in origin. 
The inferred physical 
properties of the nebula and point source (i.e. $\tau$, V$_{\rm{NS}}$, 
\.E, P, B) support our hypothesis that \ic\ and the pulsar were produced 
approximately 30,000 years ago in a core collapse supernova event.

Despite the evidence in favor of a real physical association between
\ic\ and the compact source within the cometary nebula, there are a
few puzzles that remain to be explained. As mentioned previously, one might expect the
``tail'' of the cometary nebula to point back to the origin of the SNR
and the pulsar. However, the geometric center of \ic\ lies almost due
north of the compact object, and not to the northeast as suggested by the
direction of the cometary tail.  Given the complex kinematics and
distribution of the molecular and neutral gas in the vicinity of \ic\ 
(\cite{gh79}; van Dishoeck {\it et al.} 1993) it may be overly na\"ive
to ascribe the blast center to the geometric center.  A final
problem concerns the apparent absence of X-ray and radio pulsations
from the \ic\ point source. K97 carried out a pulse search using {\it
  ROSAT} and {\it ASCA} data, while Kaspi et al.~(1996) have searched
for radio pulsations. In our view, these null results present no
immediate problem for our young neutron star hypothesis since pulsed X-rays
from Vela-like pulsars have proven difficult to detect due to their
soft spectra and contamination from the surrounding nebular emission
(e.g. \cite{bt97}).  Furthermore, it is entirely possible that the radio beam 
may not intersect our line of sight (\cite{llc98}).

\acknowledgments

We thank D. J. Thompson, B. M. Gaensler, R. Petre, P. Slane, R. A.
Chevalier, R.A. Fesen, and J. Kolena for useful discussions, and 
referee Randall Smith for his thoughtful comments.  We also thank 
G. E. Allen, K. A. Arnaud, J-H. Rho, N. Evans, and M. Johnson for 
help with data analysis.  This research was funded by NASA grant 
NRA 97-0SS-14.

\end{document}